\documentclass[12pt]{elsarticle}
\usepackage{refmerge}
\usepackage{epsfig}
\usepackage[percent]{overpic}

\def\Ecm {\ensuremath{\rm E_{\rm c.m.}}}

\def\epem {\ensuremath{e^+ e^-}}
\def\pipi {\ensuremath{\pi^+\pi^-}}
\def\ppbar {\ensuremath{p\bar p}}
\def\nnbar {\ensuremath{n\bar n}}
\def\NNbar {\ensuremath{N\bar N}}

\begin{document}
\date{\today}

\title{\bf{ \boldmath
Observation of a fine structure in $e^+e^-\to hadrons$ production at the 
nucleon-antinucleon threshold
}}

\author[adr1,adr2]{R.R.~Akhmetshin}
\author[adr1,adr2]{A.N.~Amirkhanov}
\author[adr1,adr2]{A.V.~Anisenkov}
\author[adr1,adr2]{V.M.~Aulchenko}
\author[adr1]{V.Sh.~Banzarov}
\author[adr1]{N.S.~Bashtovoy}
\author[adr1,adr2]{D.E.~Berkaev}
\author[adr1,adr2]{A.E.~Bondar}
\author[adr1]{A.V.~Bragin}
\author[adr1,adr2,adr5]{S.I.~Eidelman}
\author[adr1,adr2]{D.A.~Epifanov}
\author[adr1,adr2,adr3]{L.B.~Epshteyn}
\author[adr1,adr2]{A.L.~Erofeev}
\author[adr1,adr2]{G.V.~Fedotovich}
\author[adr1,adr2]{S.E.~Gayazov}
\author[adr1,adr2]{A.A.~Grebenuk}
\author[adr1,adr2]{S.S.~Gribanov}
\author[adr1,adr2,adr3]{D.N.~Grigoriev}
\author[adr1,adr2]{F.V.~Ignatov}
\author[adr1,adr2]{V.L.~Ivanov}
\author[adr1]{S.V.~Karpov}
\author[adr1,adr2]{V.F.~Kazanin}
\author[adr1,adr2]{I.A.~Koop}
\author[adr1]{A.N.~Kirpotin}
\author[adr1,adr2]{A.A.~Korobov}
\author[adr1,adr3]{A.N.~Kozyrev}
\author[adr1,adr2]{E.A.~Kozyrev}
\author[adr1,adr2]{P.P.~Krokovny}
\author[adr1,adr2]{A.E.~Kuzmenko}
\author[adr1,adr2]{A.S.~Kuzmin}
\author[adr1,adr2]{I.B.~Logashenko}
\author[adr1,adr2]{P.A.~Lukin}
\author[adr1]{K.Yu.~Mikhailov}
\author[adr1]{V.S.~Okhapkin}
\author[adr1]{A.V.~Otboev}
\author[adr1]{Yu.N.~Pestov}
\author[adr1,adr2]{A.S.~Popov}
\author[adr1,adr2]{G.P.~Razuvaev}
\author[adr1]{A.A.~Ruban}
\author[adr1]{N.M.~Ryskulov}
\author[adr1,adr2]{A.E.~Ryzhenenkov}
\author[adr1]{A.I.~Senchenko}
\author[adr1]{Yu.M.~Shatunov}
\author[adr1]{P.Yu.~Shatunov}
\author[adr1,adr2]{V.E.~Shebalin}
\author[adr1,adr2]{D.N.~Shemyakin}
\author[adr1,adr2]{B.A.~Shwartz}
\author[adr1,adr2]{D.B.~Shwartz}
\author[adr1,adr4]{A.L.~Sibidanov}
\author[adr1,adr2]{E.P.~Solodov\fnref{tnot}}
\author[adr1,adr2]{A.A.~Talyshev}
\author[adr1]{V.M.~Titov}
\author[adr1,adr2]{S.S.~Tolmachev}
\author[adr1]{A.I.~Vorobiov}
\author[adr1]{I.M.~Zemlyansky}
\author[adr1,adr2]{Yu.V.~Yudin}

\address[adr1]{Budker Institute of Nuclear Physics, SB RAS, 
Novosibirsk, 630090, Russia}
\address[adr2]{Novosibirsk State University, Novosibirsk, 630090, Russia}
\address[adr3]{Novosibirsk State Technical University, 
Novosibirsk, 630092, Russia}
\address[adr4]{University of Victoria, Victoria, BC, Canada V8W 3P6}
\address[adr5]{Lebedev Physical Institute, RAS, Moscow, 119333, Russia}
\fntext[tnot]{Corresponding author: solodov@inp.nsk.su}


%
\vspace{0.7cm}
\begin{abstract}
\hspace*{\parindent}
A study of  hadron production at the nucleon-antinucleon threshold has been
performed with the CMD-3 detector at the VEPP-2000  $e^+e^-$ collider. 
The very fast rise with about 1 MeV width has been observed in the $e^+e^- \to p\bar p$ cross section.
A sharp drop in the $e^+e^- \to 3(\pi^+\pi^-)$ cross section has been confirmed and found
to have a less than 2 MeV width, 
in agreement with the observed fast rise of the $e^+e^- \to p\bar p$ cross
section. For the first time a similar sharp drop is demonstrated in the
$e^+e^- \to K^+K^-\pi^+\pi^-$ cross section. The behavior of the
$e^+e^- \to 3(\pi^+\pi^-),~ K^+K^-\pi^+\pi^-$ cross sections cannot be explained
by an interference of any resonance amplitude with continuum, therefore this
phenomenon cannot be due to a narrow near-threshold resonance.  No such
structure has been observed in the $e^+e^- \to 2(\pi^+\pi^-)$ cross section.
\end{abstract}

\maketitle

\baselineskip=17pt
\section{ \boldmath Introduction}
\hspace*{\parindent}
Production of six pions in \epem annihilation, studied at 
DM2~\cite{6pidm2,6pith1,dbase}, 
showed a ``dip'' in the cross section at about 1.9 GeV, confirmed later by 
the Fermilab E687 experiment in photoproduction~\cite{focus, focus1}, 
and with a much larger effective integrated luminosity at BaBar~\cite{isr6pi}
using initial-state radiation (ISR).
Even earlier, a narrow structure near the  proton-antiproton threshold 
was also observed in the total cross section of $e^+e^-$ annihilation 
into hadrons in the FENICE experiment~\cite{fenice}.
A measurement of the CMD-3 Collaboration~\cite{cmd6pi} confirmed these 
observations and demonstrated that the drop in the $e^+e^- \to 3(\pi^+\pi^-)$ 
cross section occurred in the narrow energy range of less than 10 MeV width.  
The origin of the ``dip'' remains unclear, but one of the 
explanations suggests the presence of a below-threshold proton-antiproton
($p\bar p$) resonance~\cite{ppbartheory}.
 Alternatively, according to Refs.~\cite{han,Milst1,Milst2,Milst3} 
the ``dip'' is due to the strong interaction in virtual  nucleon-antinucleon 
($N\bar N$) production, and is related to the fast rise of the 
$e^+e^- \to N\bar N$ cross section and $N\bar N$ annihilation to hadrons. 
This hypothesis is supported by the fast increase of
the $p\bar p$~\cite{isrppbar, cmdppbar} and $n\bar n$~\cite{sndnnbar} 
form factors near threshold, and explains a similar drop in the 
$\eta'(958)\pi^+\pi^-$ spectrum, observed by the BES-III 
Collaboration in the $J/\psi \to \eta'(958)\pi^+\pi^-\gamma$ 
decay~\cite{etaprimpipi}. The authors of Ref.~\cite{han}
consider the two-step process $e^+e^- \to N\bar N \to$ multipions
and evaluate the total reaction amplitude for various intermediate mechanisms
of the  $e^+e^- \to 5\pi,~6\pi$ reactions. In 
Refs.~\cite{Milst1,Milst2,Milst3} the authors 
go even further taking into account the proton-neutron mass difference 
and $\bar{p}p$ Coulomb interaction. 

However, the mass-energy resolution of the previous experiments does not allow 
a study of the fine structure of the ``dip'' or the rise of the 
$\epem\to\NNbar$ cross section. Therefore we decided to repeat a scan
of this energy range with a larger data sample and a fine step in an
attempt to measure the width of the dip.    
In this paper we present the analysis of 
50 pb$^{-1}$ of  integrated luminosity collected with the CMD-3 
detector~\cite{sndcmd3} at 29 c.m. energy
points  at  the VEPP-2000 collider with the upgraded injection 
complex~\cite{vepp1,vepp2,vepp3,vepp4}.
While the data have been collected
in the 1.5--2.0 GeV center-of-mass energy (\Ecm) range, the scope of this paper 
is a detailed study of the \NNbar~ threshold region.  
The scan of the \NNbar-threshold energy range was performed with a fine step, 
corresponding to the c.m. energy spread.  
The beam energy and energy spread have been monitored by 
the back-scattering-laser-light system~\cite{laser1,laser2}, providing
an absolute energy measurement with better than 0.1 MeV uncertainty in every single measurement. 
During data taking the \Ecm~ variations around a central value did not exceed 
0.1 MeV at each energy point: this value is taken as the systematic uncertainty estimate. The energy spread, $\sigma_{\Ecm}$,  is measured to be
0.95$\pm$0.10 MeV at the \NNbar~ threshold: the added uncertainty is our
estimate of a systematic effect with a negligible contribution 
of the statistics.

The luminosity was measured using events of Bhabha scattering 
at large angles~\cite{lum}. 
\section{The $e^+e^-\to 3(\pi^+\pi^-)$ cross section}
\label{xs6pi}
\begin{center}
\begin{figure}[tbh]
\vspace{-0.2cm}
\begin{overpic}
[width=1.\textwidth]{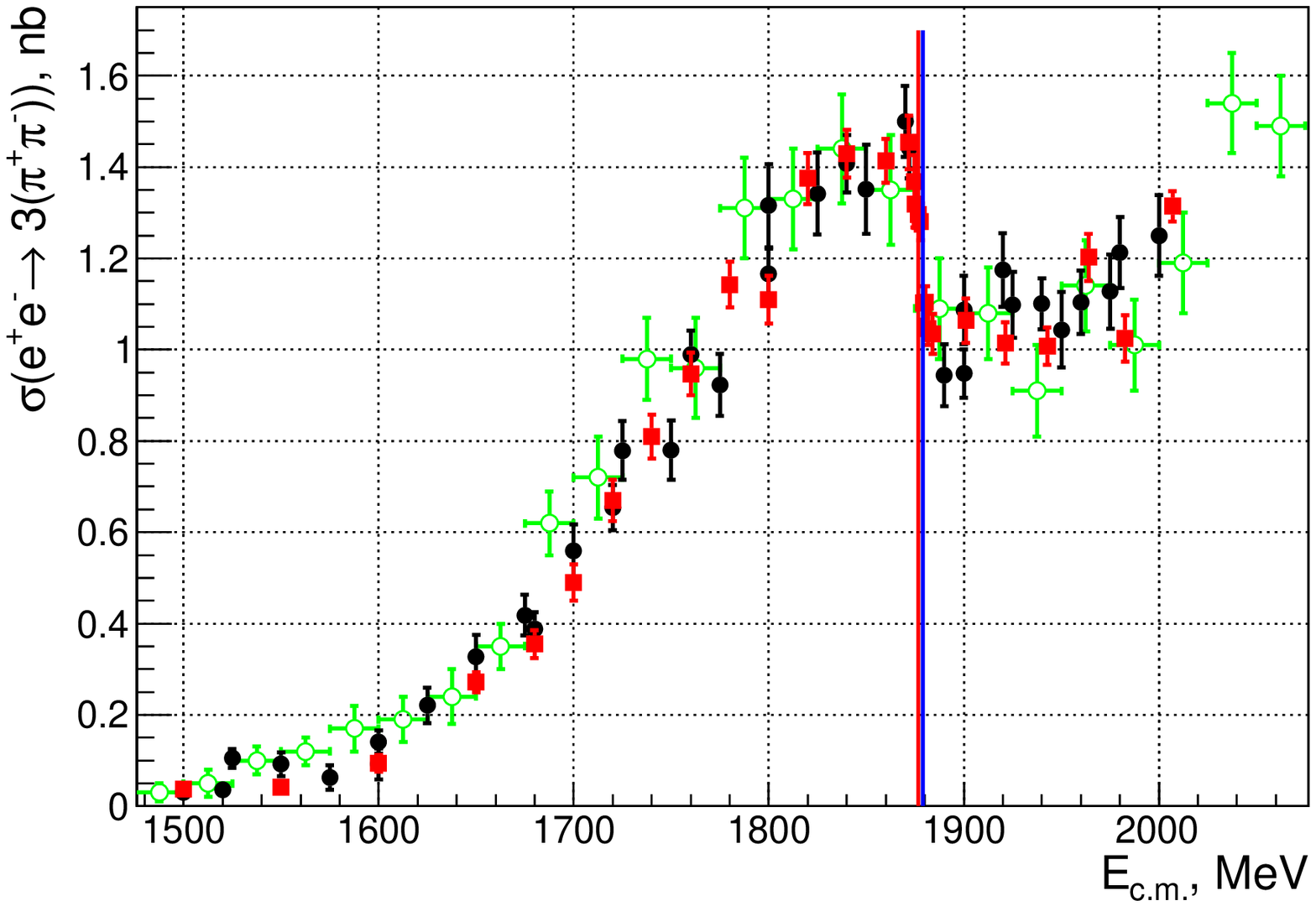}
 \put (11,37) {\includegraphics[width=0.34\textwidth]{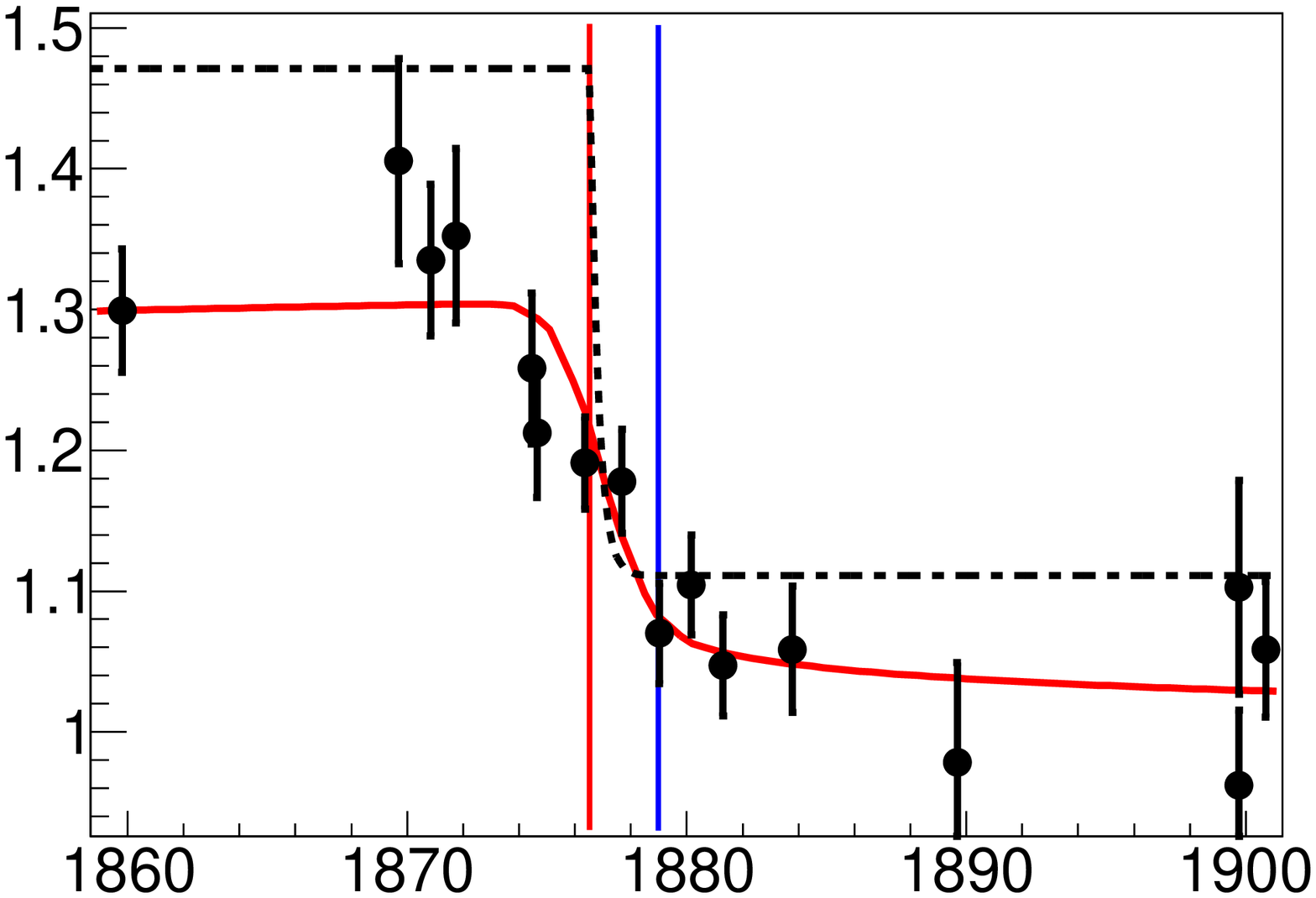}}
\end{overpic}
\vspace{-0.5cm}
\caption
{
The $e^+e^-\to 3(\pi^+\pi^-)$ Born cross section measured with the CMD-3
detector in the 2017 run (squares). The results of the previous  
CMD-3 measurement~\cite{cmd6pi} are shown by dots and those of 
BaBar~\cite{isr6pi} by open circles. The inset shows the visible cross
section with the fit 
result of a step-like function, shown by dot-dashed curve, convoluted with the resolution and radiative effects as 
described in the text. The vertical lines show   the
$\NNbar$ thresholds. Here and throughout the paper all errors in the figures 
are statistical only.
}
\label{6picross}
\end{figure}
\end{center}
\hspace*{\parindent}
The analysis of the $e^+e^-\to 3(\pi^+\pi^-)$ process was described in detail
in Ref.~\cite{cmd6pi}. For the new data we have reproduced all steps for
selection of five and six charged tracks, and the calculation of the 
efficiency and radiative corrections.   As in Ref.~\cite{cmd6pi}, we
have a background-free sample of the six-track signal events, and use 
the ratio of the five- and six-track events to correct the efficiency.
With the new data sample, the number 
of signal events with six charged tracks increased to 10155 (compared to 
2887 events in the previous analysis) and that with one missing track to 
17822 (5069) events. 
The cross section obtained from the new data is shown in Fig.~\ref{6picross} 
by squares, while the BaBar~\cite{isr6pi} and previous CMD-3~\cite{cmd6pi} 
data are shown by open and closed circles, respectively. Our 
previous result is confirmed with better statistical accuracy, while a 
systematic uncertainty is estimated at the same 6\% level, mostly dominated by the uncertainties in the efficiency and background estimate.  The ``dip'' at 
the $\NNbar$ threshold is also confirmed and is studied in more detail 
(see below).
\section{The $e^+e^-\to K^+K^-\pi^+\pi^-$ cross section}
\label{xs2k2pi}
\begin{center}
\hspace{-1.0cm}
\begin{figure}[tbh]
\vspace{-0.2cm}
\begin{overpic}
[width=1.\textwidth]{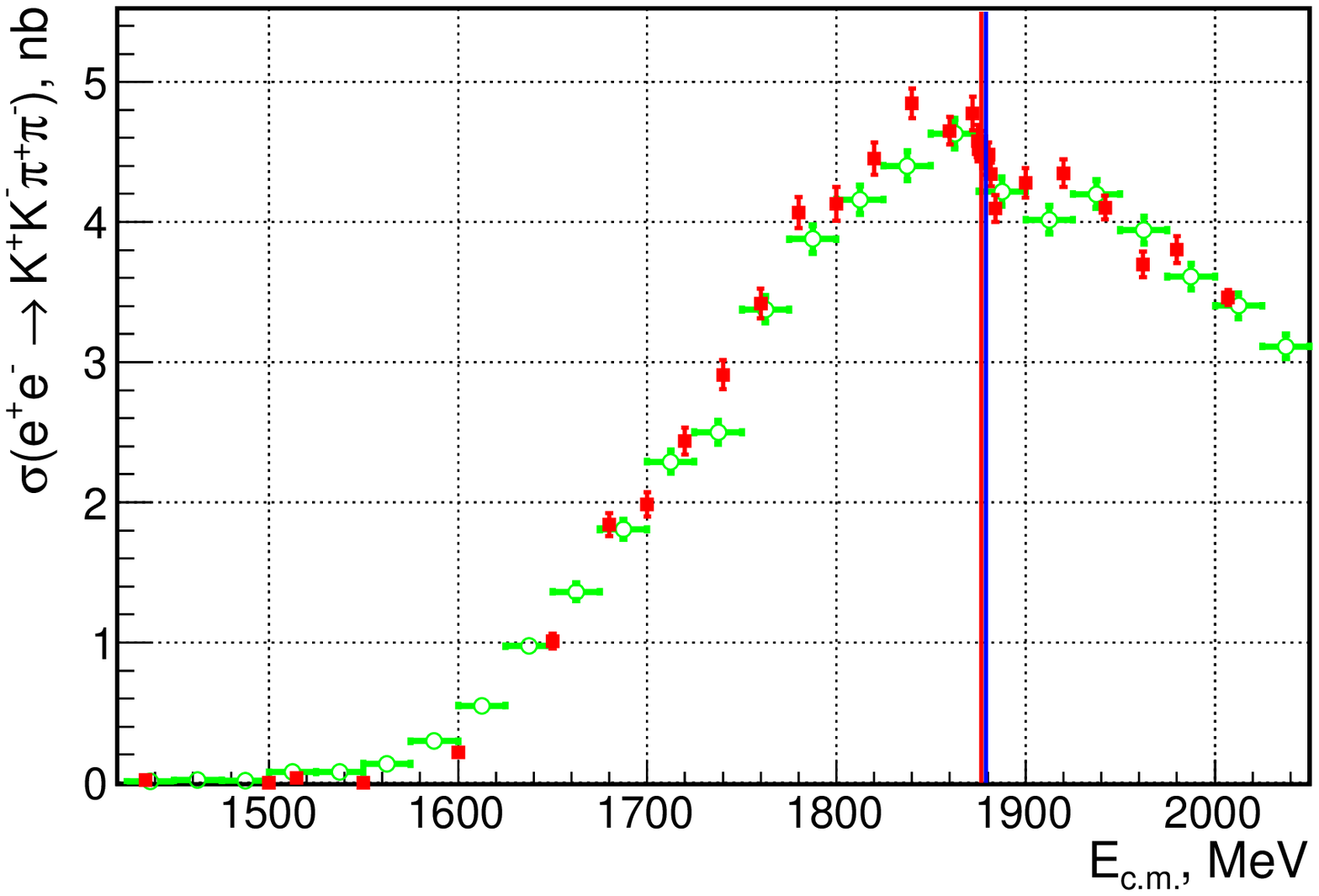}
 \put (13,37) {\includegraphics[width=0.34\textwidth]{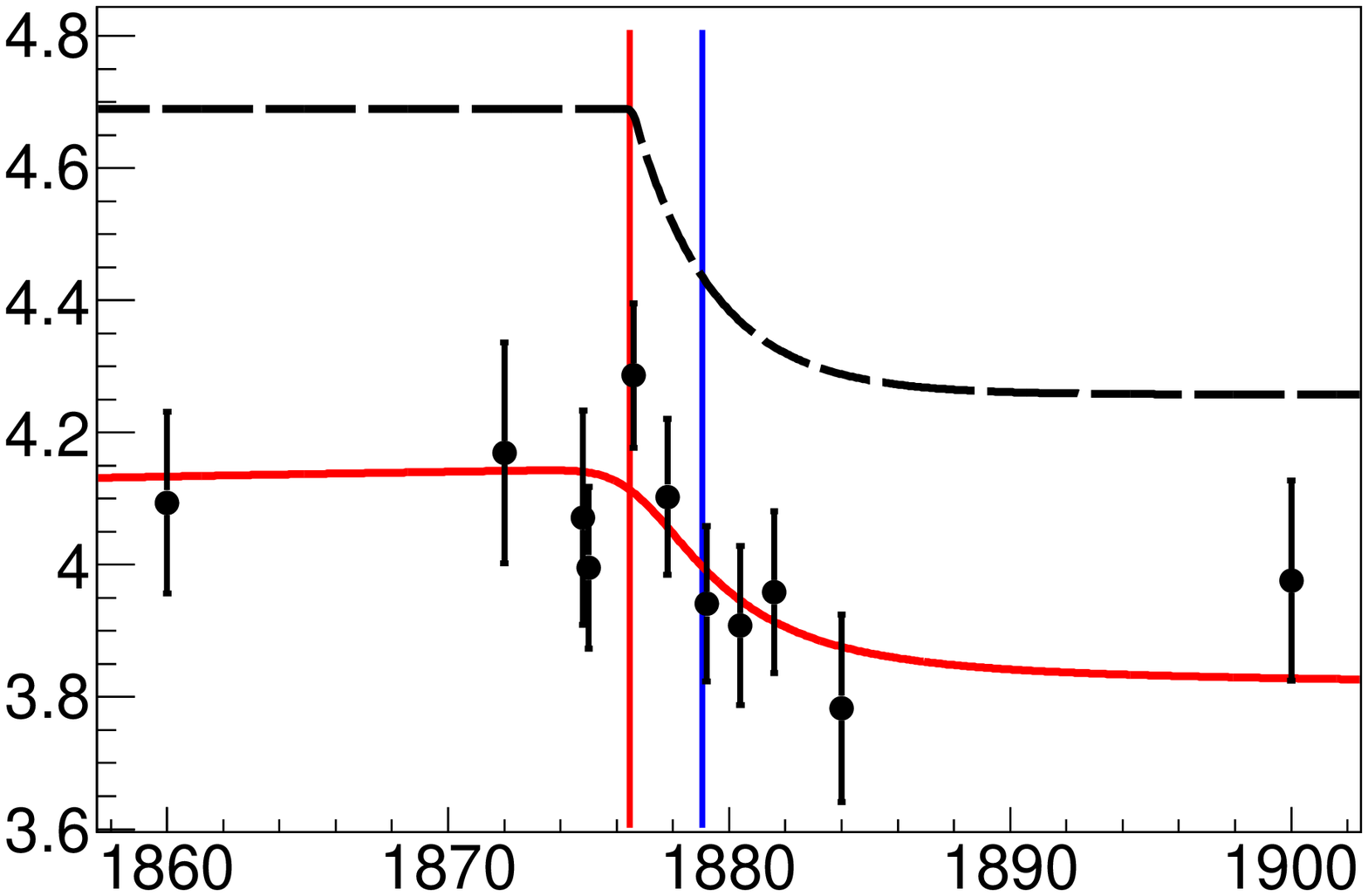}}
\end{overpic}
\vspace{-0.8cm}
\caption
{
The $e^+e^-\to K^+K^-\pi^+\pi^-$ Born cross section measured with the CMD-3
detector in the 2017 run (squares). The results of the  
BaBar~\cite{isr2k2pi} measurements are shown by open circles. The inset shows 
the visible cross section with the fit 
result of a step-like function, shown by dashed curve, convoluted with the resolution and radiative effects as 
described in the text. The
vertical lines show   the $\NNbar$ thresholds.
}
\label{2k2picross}
\end{figure}
\end{center}
\hspace*{\parindent}
The analysis of the $e^+e^-\to K^+K^-\pi^+\pi^-$ process was described in detail
in Ref.~\cite{cmd2k2pi}. For the new data we have reproduced all steps for
selection of four charged tracks, pion-kaon separation procedure, and the 
calculation of the efficiency and radiative corrections. 
 A specially designed likelihood function  is used to separate kaons and
pions. In this analysis we use events with 
exactly four charged tracks which have practically no background. 
In contrast to the previous analysis, the
events with one missing kaon or events with a missing pion are not used 
to reduce the uncertainty in the background subtraction.
Nevertheless, the same overall statistical accuracy is achieved since the 
scan around the \NNbar~ threshold is performed with large integrated
luminosity that allows us to select about 1500 signal events 
per energy point.
The cross section obtained from the new data is shown in Fig.~\ref{2k2picross} 
by squares, while the BaBar~\cite{isr2k2pi}  
data are shown by open circles. Our previous result is confirmed with better 
statistical accuracy, while a systematic uncertainty remains at the same 6\% 
level, dominated by that in the efficiency estimate. 
Evidence for the  ``dip'' at the $\NNbar$ threshold is obtained for the first time 
in this channel and is studied in more detail below.
\section{The $\epem\to\ppbar$ cross section at the $\NNbar$ threshold}
\label{xsppbar}
\begin{center}
\begin{figure}[tbh]
\begin{overpic}
[width=1.\textwidth]{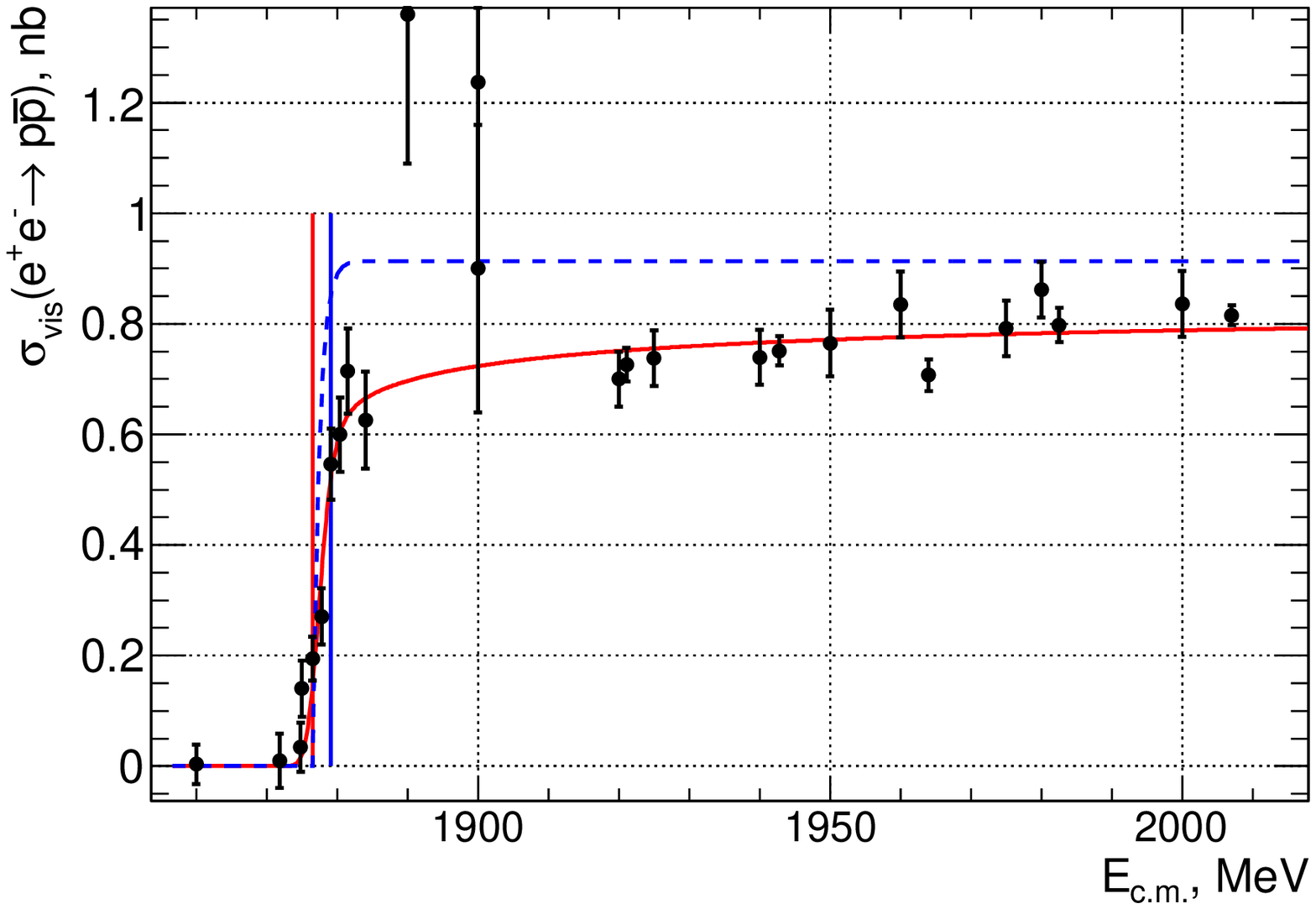}
 \put (55,12) {\includegraphics[width=0.39\textwidth]{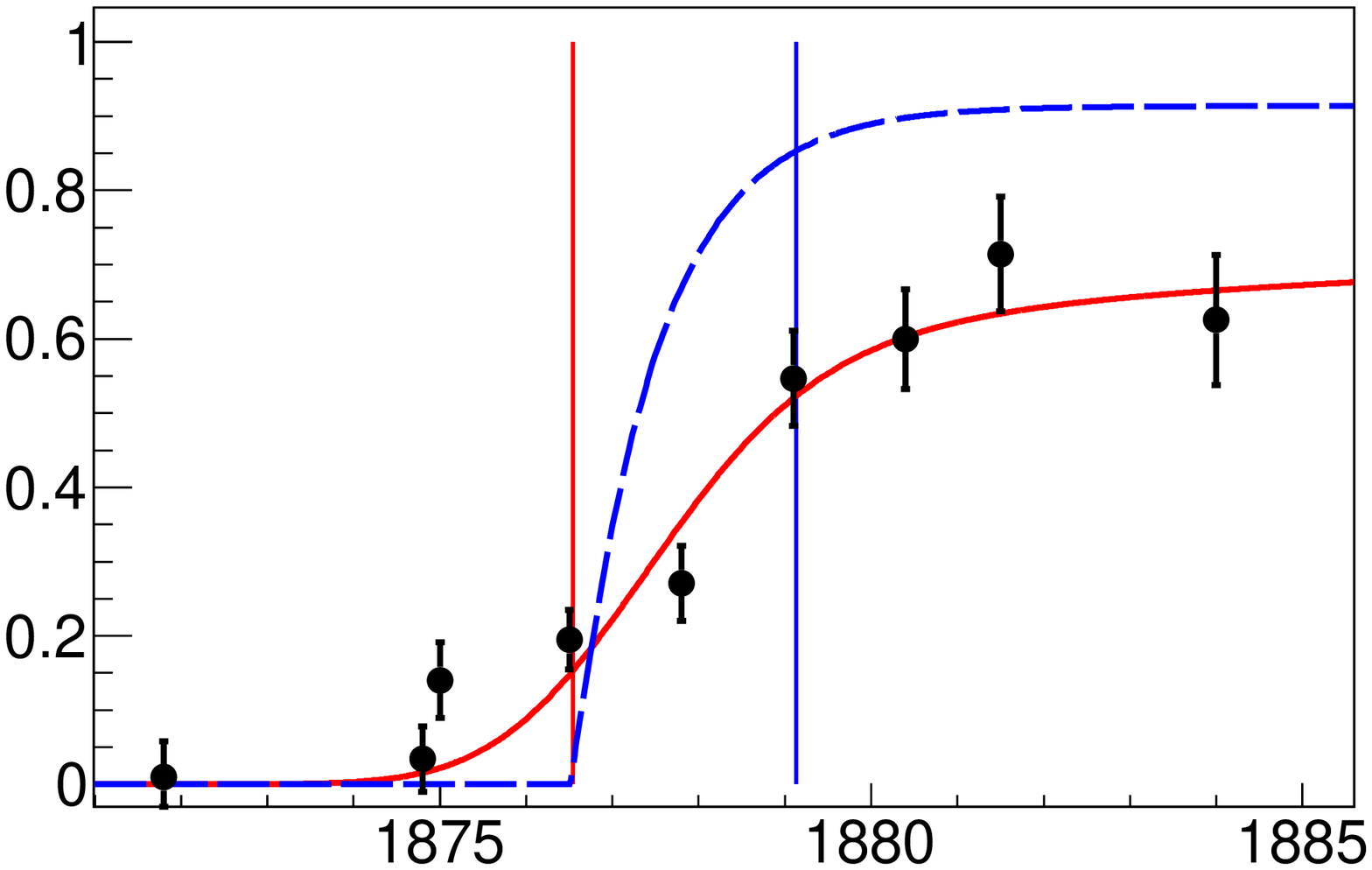}}
\end{overpic}
\vspace{-0.8cm}
\caption
{
The $\epem\to\ppbar$ visible cross section measured with the CMD-3
detector. The solid curve shows the result of the fit to an exponentially 
saturated function of Eq.(3) (shown by dashed curve, $\rm E_{\rm thr}$ is fixed at 1876.54 MeV) convolved with the 0.95 MeV energy spread and radiation functions.
The vertical lines show the $\ppbar$ and $\nnbar$ thresholds. The inset shows 
the expanded view of the visible cross section.
}
\label{ppbarthres}
\end{figure}
\end{center}

\hspace*{\parindent}
The analysis procedure is described in our previous 
publication~\cite{cmdppbar}. At the energies near threshold, for
$\Ecm < 1900$ MeV,  
protons and antiprotons from the reaction $\epem\to\ppbar$ stop in the 
material of the beam pipe because of very low momentum. To select such 
events, we look for the products of antiproton annihilation with more than 
two charged tracks coming from the aluminum beam pipe. Comparison of the 
calorimeter response for such events below and above the $\NNbar$ threshold 
yields the number of $\ppbar$ events. Points below the production thresholds, 
where we assume no signal from the $\epem\to\ppbar$ reaction, are used for 
background normalization and we obtain 490$\pm$30 signal events in the
energy range from the production threshold to 1900 MeV.   
Starting from \Ecm=1900 MeV, protons have enough energy to penetrate the beam 
pipe, and above this energy no annihilation of antiprotons at the beam pipe 
is observed. 
Protons and antiprotons are detected as collinear tracks with large 
specific energy losses, dE/dx, in the drift chamber (DC) of the CMD-3: we 
detect 4770 signal events. 
At each energy a visible cross section is calculated 
as the number 
of selected events divided by the detection efficiency and integrated 
luminosity. The obtained $\epem\to\ppbar$  visible cross section 
is shown in Fig.~\ref{ppbarthres}. We estimate the systematic uncertainty as about 10\%, dominated by the uncertainty in the efficiency calculation: a special study was performed to estimate data-MC difference in the reconstruction efficiency.

\section{The $\NNbar$ threshold region}
\label{xsnnbar}
The cross section in Fig.~\ref{ppbarthres} exhibits very sharp 
step-like behavior close to the $\NNbar$ threshold.  The Born cross section
cannot be obtained without taking into account  its
 smearing due to radiation of real photons by 
initial electrons and positrons, and the energy spread of the
collision energy
with $\sigma_{\Ecm} = 0.95\pm 0.10$ MeV. The visible cross section is described 
by a convolution of the radiative cross section, $\sigma_{f\gamma}(\Ecm)$, 
with the c.m. energy spread function:
\begin{eqnarray}
\label{sigborn}
\sigma_{\rm vis}(\Ecm) = \frac{1}{\sqrt{2\pi}\sigma_{\Ecm}}\int d\Ecm'
  \sigma_{\rm f\gamma}(\Ecm')\cdot
exp\left(- \frac{(\Ecm - \Ecm')^2}{2\sigma_{\Ecm}^2}\right),
\end{eqnarray}
where  $\sigma_{\rm f\gamma}(\Ecm)$ is a convolution of the Born cross section 
with the radiator function $\rm F(\Ecm, E_{\gamma})$ ~\cite{kur_fad,actis}:
\begin{eqnarray}
\label{sigvis}
\sigma_{\rm f\gamma}(\Ecm)=\int_0^{E_{\gamma}^{max}}
  dE_{\gamma}\cdot\sigma_{\rm Born}\left(\Ecm\sqrt{1 -
  E_{\gamma}/\Ecm}\right)\cdot
 F\left(\Ecm, E_{\gamma}\right),
\end{eqnarray}
where $\rm E_{\gamma}$ is the radiative photon energy, and 
 $\rm E_{\gamma}^{\rm max}$ is a maximum allowed photon energy for the reaction.

For a demonstration of very fast variation of the cross section, 
$\sigma_{\rm Born}(\Ecm)$ is described with an exponentially saturated function, 
\begin{eqnarray}
\label{fermi}
\sigma_{\rm Born}(\Ecm) = A + B\left[1 - exp\left(- \frac{(\Ecm - E_{thr})}{\sigma_{thr}}\right)\right],
\end{eqnarray}
where $\rm E_{\rm thr}$ ($\Ecm > E_{\rm thr}$) and $\rm \sigma_{\rm thr}$ are the 
energy threshold and a variation scale of the Born cross section, 
respectively. The values of $\rm A$ and $\rm A +\rm B$ ($\rm B = 0$ for $\Ecm < E_{\rm thr}$) give the asymptotic values of the cross section below 
and above the $\ppbar$ threshold.  

 First, the $\epem\to\ppbar$ visible cross section is fit to Eq.~\ref{fermi} with all parameters floating 
except the $\rm A$ value fixed at zero, assuming no signal below the threshold.
The fit yields $\rm E_{\rm thr} = 1877.1\pm0.2$ MeV, 
consistent with the $\ppbar$  production threshold within  
uncertainties in the energy measurement, and  
$\rm \sigma_{\rm thr} = 0.18 \pm 0.27$ MeV.
Since no $\ppbar$ events are expected below the threshold,  
$\rm E_{\rm thr}$ is fixed at 1876.54 MeV (the 
doubled proton mass), and the fit yields $\rm \sigma_{\rm thr} = 0.76 \pm 0.28$ 
MeV. In both cases the $\rm \sigma_{\rm thr}$ value and its uncertainty are 
smaller than the 
energy difference between the neutron and proton production thresholds.
Figure~\ref{ppbarthres}
shows the visible $\epem\to\ppbar$ cross section with the fit result (solid curve) when  $\rm E_{\rm thr}$ in Born cross section (dashed curve) is fixed at 1876.54 MeV.
Lines show the $\ppbar$ and $\nnbar$
threshold positions. An expanded view of the visible $\epem\to\ppbar$
cross section around the $\NNbar$ threshold is shown in the inset in 
Fig.~\ref{ppbarthres}.

\begin{table}
\caption{Results of the fit to the exponentially rising function. Only statistical uncertainties are shown.}
\label{tab}
\begin{tabular}{|c|c|c|c|c|c|}
\hline
Reac. & A, nb & B, nb & E$_{\rm thr}$, MeV &  $\rm \sigma_{\rm
  thr}$, MeV & $\chi^2/ndf $ \\
\hline
\ppbar & 0~--~fxd & 0.91$\pm$0.02 & 1877.1$\pm$0.2 &
                                                      0.18$\pm$0.27&29/26\\
\ppbar & 0~--~fxd & 0.91$\pm$0.02 & 1876.54-fxd &
                                                      0.76$\pm$0.28&31/27 \\
$6\pi$ &1.55$\pm$0.02&--0.42$\pm$0.03&1875.8$\pm$0.2&0.18$\pm$0.67&17/20\\ 

$6\pi$ &1.54$\pm$0.02&--0.41$\pm$0.03&1876.54--fxd&0.0$\pm$2.5&18/21\\

$2K2\pi$ &4.69$\pm$0.08&--0.44$\pm$0.12&1878.8$\pm$0.2&0.35$\pm$2.69&7/10\\ 

$2K2\pi$ &4.70$\pm$0.08&--0.45$\pm$0.12&1876.54--fxd&2.36$\pm$2.01&8/11\\
\hline
\end{tabular}
\end{table}

Similarly, the $e^+e^-\to 3(\pi^+\pi^-)$ visible cross section is fit
to the above functions with all parameters floating  
in the energy range \Ecm = 1834--1944 MeV, where the cross section can be 
considered relatively flat. The fit yields $\rm E_{thr} = 1875.8 \pm 0.2$ MeV, and 
$\rm \sigma_{thr} = 0.18 \pm 0.67$ MeV.  The fit with fixed 
$\rm E_{thr} = 1876.54$ MeV  yields $\rm \sigma_{thr} = 0.0 \pm 2.5$ MeV, 
 with a good $\chi^2/ndf = 18/21$ value: our statistical accuracy and energy spread 
allow a drop with 
a zero width.  The result of the latter fit is shown as an inset in
Fig.~\ref{6picross} by a solid line, while a dot-dashed line shows the 
Born cross section.  
The obtained $\rm \sigma_{thr}$ value is consistent with that obtained for 
the $\epem\to\ppbar$ reaction. Variation of the ``flat'' region in reasonable 
scale changes the $\rm B$ value, but has small influence on 
the ``dip'' parameters.

Then we fit the $e^+e^-\to K^+K^-\pi^+\pi^-$ visible cross
section to the above functions with all parameters floating in the
energy range \Ecm = 1850--1970 MeV, where the cross section can be  
considered relatively flat. The fit yields $\rm E_{thr} = 1878.8 \pm 0.2$ MeV:
the value is close to the \nnbar~threshold.  The obtained value
$\rm \sigma_{thr} = 0.35 \pm 2.69$ MeV indicates that the observed effect 
is dominated by the statistical uncertainty, and is consistent with 
a zero-width drop in the Born cross section.  
The fit with fixed $\rm E_{thr} = 1876.54$ MeV  yields 
$\rm \sigma_{thr} = 2.36 \pm 2.01$ MeV, 
 with a good $\chi^2/ndf = 8/11$ value.  The result of the
latter fit is shown as an inset in Fig.~\ref{2k2picross} by a solid line,
while a dashed line shows the Born cross section.  

The results of the fit are summarized in Table~\ref{tab}, and  demonstrate 
that the observed behavior of the cross sections has 
similar origin, and the ``dip'' in the hadronic cross section can be 
interpreted as due to opening of the direct production of the $\NNbar$ channel.
Note that when $\rm E_{thr}$ is floating, the obtained value in case of $3(\pipi)$ is close to the \ppbar~ threshold energy, while for the $K^+K^-\pi^+\pi^-$ channel this value is consistent with the \nnbar~ threshold (1879.13 MeV). 

We perform a simultaneous fit of all three channels with common 
$\rm E_{thr}$ and $\rm \sigma_{thr}$ values, and the fit yields
$1876.87\pm0.10\mp0.11$ MeV and  $0.31\pm0.25\mp0.15$ MeV, respectively, with
$\chi^2/ndf = 66/(67-7)$ value. The second uncertainty is systematic
and anticorrelated with the systematic uncertainty in the energy spread
$0.95\pm0.10$ MeV.

Unfortunately, the accelerator-induced 
energy spread and relatively low statistical accuracy do not allow us
to directly observe a possible structure of this rise (drop) due to the
proton-neutron interaction, which could be expected in the studied 
reactions. 

\begin{center}
\begin{figure}[tbh]
\vspace{-0.2cm}
\begin{overpic}
[width=1.0\textwidth]{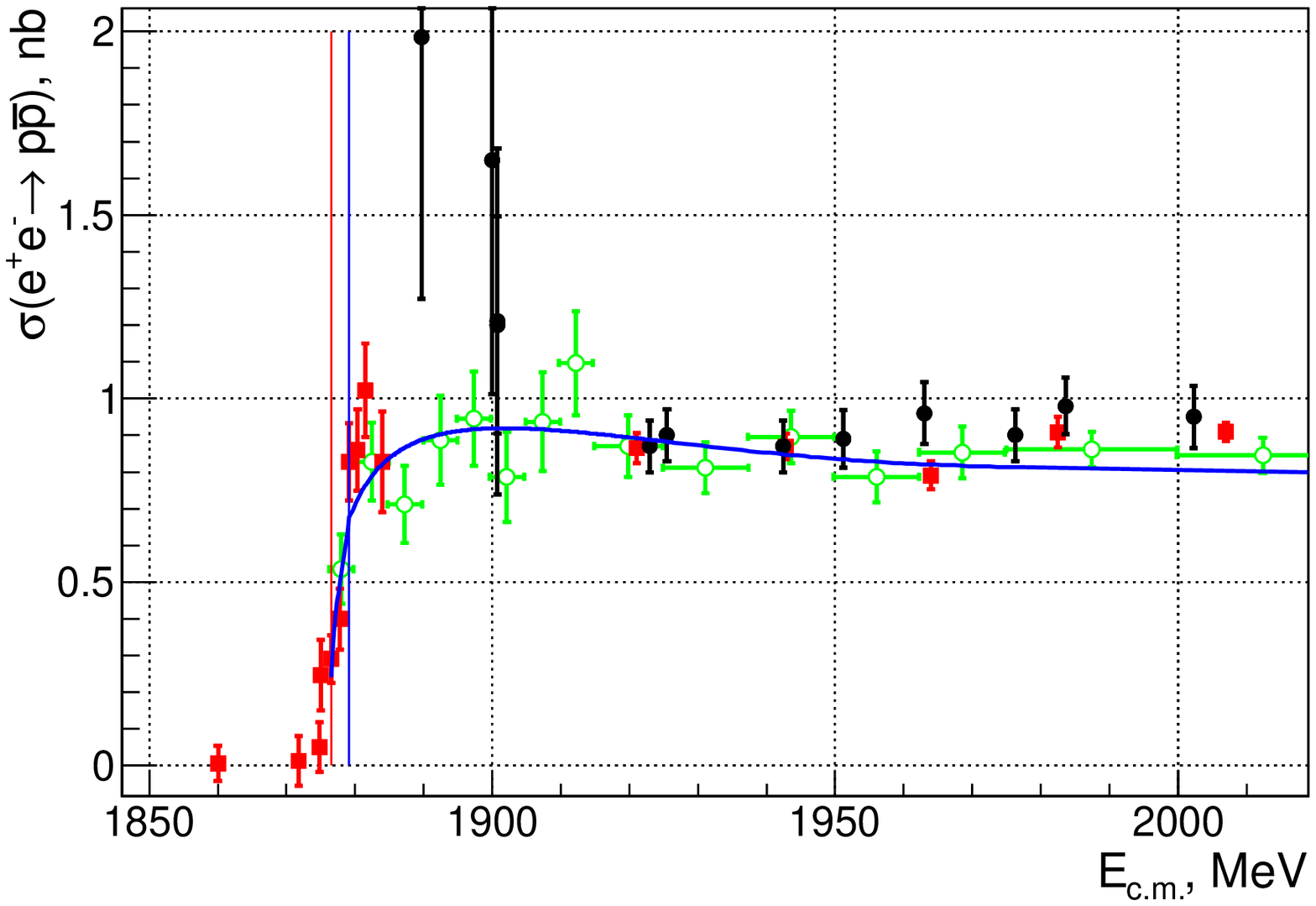}
 \put (50,35) {\includegraphics[width=0.37\textwidth]{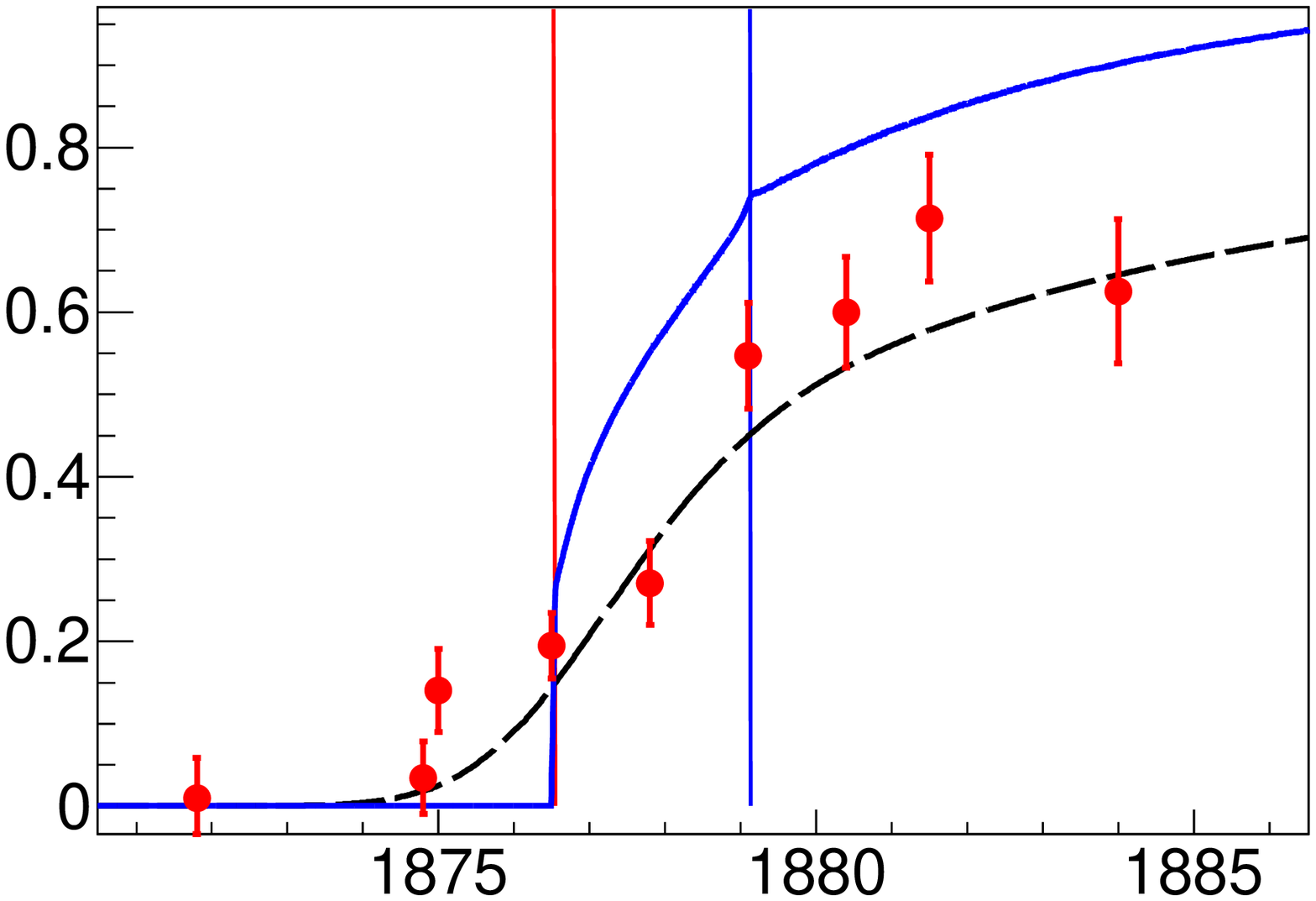}}
\end{overpic}
\vspace{-0.6cm}
\caption
{
The $\epem\to\ppbar$ Born cross section measured with CMD-3 (dots ~\cite{cmdppbar} and squares) and BaBar (open circles).
The solid curve shows the result of the prediction 
from Refs.~\cite{Milst1,Milst2,Milst3}.
The inset shows the expanded view of the theoretical function for the Born  (solid line), and for the visible cross section after experimental effects (dashed line) in comparison with the CMD-3 data.
The vertical lines show the \ppbar~and \nnbar~thresholds.
}
\label{ppbar_all}
\end{figure}
\end{center}

In a recently published paper~\cite{Milst3}, the authors use the optical 
potential and experimental data of the nucleon interactions to make a prediction of the \ppbar~and \nnbar~cross section behavior at very small energies above the production thresholds.  The calculated theoretical $\epem\to\ppbar$ Born cross section is shown in
Fig.~\ref{ppbar_all} by solid curve and is in good agreement with available data. 
But for a comparison of the theoretical curve and data at very small deviations from the threshold, energy spread and radiative effects must be taken into account. The result of 
this convolution for the theoretical function is shown in the inset (dashed curve), and also is in good agreement with our visible 
cross section. Note, the suggested model of the final-state interaction of 
a very slow $\NNbar$ pair predicts a nonzero cross section at the 
\ppbar~threshold due to the Coulomb interaction, but experimental effects 
and limited accuracy do not allow us to prove that.

\section{The $\epem\to 2(\pipi)$ cross section at the $\NNbar$ threshold}
\label{4pi}

As suggested in Ref.~\cite{Milst3}, the total hadronic cross section is  
strongly affected by virtual production and annihilation of the $\NNbar$ pairs. 
The calculation predicts a 7 nb ``bump'' in the total cross section, 
which is about 40 nb at this energy, and should be seen in all 
$\epem\to hadrons$ final states. A naive expectation suggests that the 
effect could be proportional to the  probability of \ppbar~annihilation 
into the studied final state.
\begin{center}
\begin{figure}[tbh]
\vspace{-0.2cm}
\includegraphics[width=1.\textwidth]{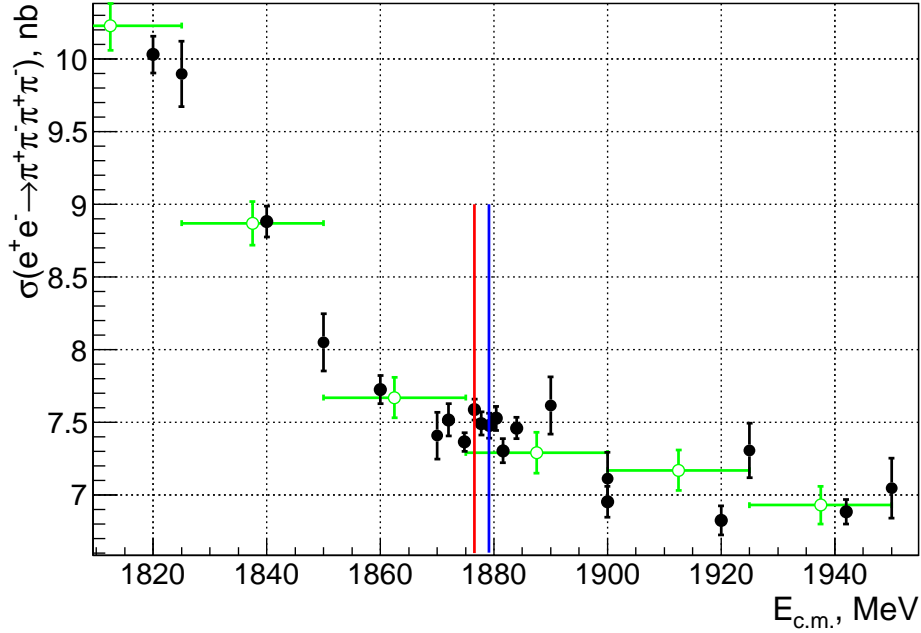}
\vspace{-0.5cm}
\caption
{
The $e^+e^-\to 2(\pi^+\pi^-)$ cross section measured with CMD-3 (dots) and 
BaBar~\cite{isr4pi} (open circles). The vertical lines show the \ppbar~and 
\nnbar~thresholds.
}
\label{xs4pi}
\end{figure}
\end{center}
To test that, we analyze data at the $\NNbar$ threshold by selecting events 
for the reaction $\epem\to 2(\pipi)$ according to the procedure described in 
Ref.~\cite{cmd4pi}, and show the obtained cross section in Fig.~\ref{xs4pi} 
together with the most precise measurement by BaBar~\cite{isr4pi}. 
While the overall systematic uncertainties on the cross section are still 
under investigation, the relative point-to-point errors do not exceed 
0.1-0.2 nb. No structure exceeding the level of 0.1 nb is observed 
at the $\NNbar$ threshold 
in either measurement. According to Ref.~\cite{ppbarreview}, the 
probability of \ppbar~annihilation (with isospin one) to four charged pions is 
about 14\%, while for six charged pions it is about 6\%. If a cross section 
drop in the hadronic channel is related to virtual 
$\NNbar$ annihilation~\cite{Milst3}, for four-pion production one could expect 
an about 0.5--0.8 nb drop in the cross section, which is not supported by our 
data. Note that according to Ref.~\cite{ppbarreview} the probability of
$\NNbar$ annihilation to the $K^+K^-\pi^+\pi^-$ final state is much
lower than that for six- or four-pion states, and observation of the
``dip'' in this channel indicates a complicated production dynamics.
\section*{ \boldmath Conclusion}
\hspace*{\parindent}
Using the improved performance of VEPP-2000, the scan of the $\epem$ c.m. 
energy in the 1680 -- 2007 MeV range has been carried out. A detailed study of
the $\NNbar$ threshold region confirms a fast drop (rise) in the  
$e^+e^-\to 3(\pi^+\pi^-)$ ($\epem\to\ppbar$) cross section observed 
previously. For the first time a width of this structure is measured 
in the $\epem\to\ppbar$ reaction: 
the $\sigma_{\rm thr} = 0.76\pm0.28$ MeV value is smaller than the 
difference between the \ppbar~and \nnbar~production thresholds.
The energy position of the ``dip'' in the $e^+e^-\to K^+K^-\pi^+\pi^-$ 
cross sections,  observed for the first time, is consistent with 
the \nnbar~ production threshold, while that for the $e^+e^-\to 3(\pi^+\pi^-)$ 
reaction is close to the \ppbar~ threshold.
No structures in the $\epem\to 2(\pi^+\pi^-)$ cross section have been 
found at the $\NNbar$ threshold suggesting
 for a more complicated dynamics at the microscopic level, which cannot be summarized by the simple rise of the virtual nucleon-antinucleon production.

After this work was submitted to the arXiv, the paper~\cite{lichard}
appeared in which the author analyzes various final states of 
$e^+e^-$ annihilation around the \NNbar~ threshold. Based on the earlier data
with a typical c.m. energy step of 10 MeV or larger, he claims the existence 
of the $\rho(1900)$ resonance residing above the \nnbar~ threshold with a
width of about 10 MeV. In contrast, our new data show that the observed 
structure is consistent with opening \NNbar~ thresholds and has a
much smaller width.  

\subsection*{Acknowledgments}
\hspace*{\parindent}
The authors are grateful to A.~I.~Milstein for useful discussions and
help with a theoretical interpretation. 
We thank the VEPP-2000 personnel for excellent machine operation. Part of 
this work related to the photon reconstruction algorithm in the 
electromagnetic calorimeter is supported by the Russian Science Foundation 
(project \#14-50-00080). The work is partially supported by the Russian 
Foundation for Basic Research grants 16-02-00160-a,
17-02-00327-a, 17-52-50064-a and 18-32-01020. Part of this work related
to account of theshold proximity is supported by the MSHE grant   
14.W03.31.0026.

\end{document}